\documentclass[final,5p,times,twocolumn]{elsarticle}
\setcitestyle{number}
\usepackage[utf8]{inputenc}
\usepackage{bm,amsmath,amssymb}
\usepackage{graphicx}
\usepackage{subfigure}
\usepackage{color}
\usepackage{hyperref}
\usepackage{multirow}
\usepackage{soul}
\usepackage{comment}
\usepackage{slashed}

\let\oldalign\align
\def\align{\linenomath\oldalign}

\newcommand{\xs}{{\bf x}}
\newcommand{\ys}{{\bf y}}

\newcommand{\ps}{{\bf p}}

\newcommand{\ib}{{\bf i}}

\begin{document}

\title{Momentum Dependence of Heavy Quark Diffusion in a Thermal Gluonic Plasma on the Lattice} 

\author[first,second]{Harshit Pandey\corref{cor1}}

\author[first,second]{Sayantan Sharma}
\affiliation[first]{The Institute of Mathematical Sciences, Chennai 600113, India}
\affiliation[second]{Homi Bhabha National Institute, Training School Complex, Anushaktinagar, Mumbai 400094, India}
\cortext[cor1]{Corresponding author, email: harshitp@imsc.res.in}

\begin{abstract}	
We study the dynamics of a heavy quark in a thermal plasma consisting of non-perturbatively interacting soft 
momentum gluons at high temperatures, described in terms of an effective theory of QCD. Discretizing this effective 
field theory on a three-dimensional lattice, we propose a numerical strategy that allows us to simulate the dynamics of 
a heavy quark for different values of initial momenta and for a wide temperature range, higher than $480$ MeV. 
This allows us, for the first time, to extract the momentum dependence of the heavy quark drag and diffusion coefficients in a non-perturbatively interacting thermal, non-Abelian plasma. 
\end{abstract}

\maketitle

\section{Introduction}
Heavy quarks are produced at the very early stages of a heavy-ion collision event due to perturbative
hard scattering of the valence partons~\cite{Andronic:2015wma}. This allows them to explore the entire 
evolution of the medium formed in a heavy-ion collision event due to the sea gluons with non-perturbatively 
large initial occupation numbers, leading to the emergence of quark-gluon plasma (QGP) which eventually hadronizes. 
The transverse momentum $p_T$ spectrum of the produced heavy quark pairs due to heavy-ion collision exhibits 
a power-law dependence, which is very different from a thermal spectrum. The question one would like 
to ask is whether the heavy quark momentum distribution acquires a thermal spectrum, and if yes,  
on what time-scale does this kinetic equilibration occur. This question has become even more relevant 
in recent years due to the observation that the elliptic flow of prompt $D$ mesons with lower transverse 
momenta is almost as strong as pions~\cite{ALICE:2020iug}, suggesting that the kinetic equilibration 
time for charm quarks should be lower than the QGP lifetime. On the other hand, elliptic flow 
of non-prompt $J/\psi$~\cite{PHENIX:2024axj} and $D$ mesons~\cite{ALICE:2023gjj}, which are produced 
due to the decay of beauty-hadrons, is suppressed, indicative of the fact that the bottom quarks might not 
achieve kinetic equilibration.

For estimating the kinetic equilibration time for heavy quarks of mass $M$ from the fundamental theory 
i.e QCD, the quantity of interest is the heavy quark momentum diffusion coefficient $\kappa$
~\cite{Svetitsky:1987gq,Casalderrey-Solana:2006fio,Rapp:2018qla,He:2022ywp}. In the limit when $M\gg T$ 
and negligibly small velocity, where $T$ is the temperature of the QGP medium, the drag $\eta_D$ 
and diffusion coefficients are related by Einstein relation $\eta_D=\kappa/(2MT)$. 
The kinetic equilibration time-scale is simply the inverse of the drag 
coefficient. Perturbative values of $\kappa$ within the hard thermal loop effective theory of 
QCD have a very poor convergence; the next-to-leading corrections~\cite{Caron-Huot:2008dyw} are more 
than twice as large compared to leading order results~\cite{Braaten:1991jj,Braaten:1991we}, for 
strong couplings at temperatures relevant for heavy-ion colliders. Extracting $\kappa$ from the 
spectral function corresponding to the two-point correlator, within the Kubo formalism, is 
challenging as the transport peak has a width $\sim 1/M$ which is difficult to resolve~\cite{Altenkort:2023eav}. 
Instead, in the static limit, $\kappa$ can be extracted from the color electric field correlators~\cite{Laine:2009dd}
within the heavy quark effective theory, whose spectral function is linearly proportional to 
frequency $\omega$ for its infinitesimal values. $\kappa$ has been estimated using lattice 
techniques, both in quenched thermal
~\cite{Banerjee:2011ra,Francis:2011gc,Francis:2015daa,Altenkort:2020fgs,Brambilla:2020siz,Banerjee:2022gen, Brambilla:2022xbd} as well as in an overoccupied non-thermal gluon plasma~\cite{Boguslavski:2020tqz, Pandey:2023dzz}. 
Lattice calculations also performed in 2+1 flavor QCD with physical quark masses~\cite{Altenkort:2023oms} 
show significant non-perturbative contribution~\cite{Casalderrey-Solana:2006fio} due to the soft gluons. 
Going beyond the static limit involves calculating the coefficient of the $\mathcal{O}(v^2)$ term, which can 
be obtained from the two-point correlation function of color magnetic fields~\cite{Bouttefeux:2020ycy}. 
This correlation function requires non-trivial renormalization~\cite{Altenkort:2024spl,delaCruz:2024cix}. 
Nevertheless, the values of $\kappa$ calculated on the lattice at $\mathcal{O}(v^2)$ both in 
quenched~\cite{Banerjee:2022uge,Altenkort:2021ntw,Brambilla:2022xbd} 
as well as in 2+1 flavor QCD~\cite{Altenkort:2023eav,HotQCD:2025fbd} show more than $16\%$ 
difference from the calculations done only with color electric field correlators for charm quarks. 
Hence, it is important to revisit the calculation of $\kappa$ on the lattice beyond 
an order-by-order expansion about the static limit as a function of their velocity $v$, by treating 
them as Dirac fermions.

It is also important to emphasize that $\kappa$ is a function of the momentum $\mathbf{p}$ 
with which the heavy quark traverses the QGP medium. The lattice calculations reviewed till now 
can only give us $\kappa(\vert \mathbf{p}\vert =0)$ since the formalism is based on an expansion about 
the static limit for heavy quarks. Analytic results for the momentum dependence of drag and diffusion 
coefficients exist within HTL perturbation theory for $M\gg T$, in leading logarithm in $m_D/T$~\cite{Moore:2004tg}, 
where $m_D$ is the Debye mass. These coefficients have also been estimated from a classical Boltzmann 
framework for charm quarks~\cite{Svetitsky:1987gq}. The diffusion coefficient of a heavy quark can also 
be extracted from the rate of collisional energy loss calculated within HTL~\cite{Braaten:1991jj,Braaten:1991we}. 
We, for the first time, calculate $\kappa$ as the function of $\vert\mathbf{p}\vert$
and at a fixed $T$ in a thermal effective theory of the non-perturbatively interacting soft 
gluons~\cite{Bodeker:1998hm} whose momenta are less than the magnetic scale $g^2 T/\pi$, using 
ab-initio lattice techniques. This will allow us to understand how heavy quarks produced at different 
initial momenta that are not ultra-relativistic, will lose energy inside the QGP due to collisions with 
the soft gluons~\cite{Svetitsky:1987gq,vanHees:2004gq,Moore:2004tg,Mustafa:2004dr} and eventually 
attain kinetic equilibration~\cite{Svetitsky:1987gq,GolamMustafa:1997id} with the thermal plasma.

In this work, we have generalized our formalism which was previously used to extract the diffusion 
coefficient for heavy quarks with zero initial momentum in a non-thermal SU(2) plasma~\cite{Pandey:2023dzz}, 
to the case where these are produced with a net initial momentum.  We then study the dynamics of these 
heavy quarks in a thermal plasma consisting of strongly interacting soft gluons~\cite{Bodeker:1998hm}, 
on a three-dimensional spatial lattice. This allows us, for the first time, to estimate 
non-perturbatively the drag and diffusion coefficients for heavy quarks with $M\gg T$ within lattice 
QCD. The outline of this letter is as follows: In the next section, we describe our 
formalism and the observables we will be calculating, followed by a discussion of the 
specific numerical trick that we use for our problem at hand. Subsequently, we discuss 
our results, the highlight among them being a full non-perturbative determination of $\kappa(\vert \mathbf{p}\vert)$ 
for a range of initial momenta for heavy quarks. We compare our new results with the ones that 
are available in the literature from lattice QCD studies at $\vert \ps \vert=0$. We end by discussing 
the physical implications of our findings that might be relevant for heavy-ion experiments and an 
outlook for future studies in this direction.

\section{Formalism }

We consider heavy quark dynamics in a thermal non-Abelian plasma at high temperatures, 
where the hard, electric, and magnetic scales are well separated as $g^2T/\pi \ll gT \ll \pi T$.  
In such a case, the hard and the electric momentum modes of the gluons can be integrated 
out to give us an effective field theory~\cite{Bodeker:1998hm} of the non-perturbatively 
interacting soft magnetic gluons. The effective theory does not suffer from the typical 
Rayleigh-Jeans divergence that plagues a classical theory of gluons. Within this effective 
theory, the soft electric fields $ E^i(t,\mathbf{x})~,i=1,2,3$ evolve according to a Langevin 
equation,
\begin{equation}
-\partial_t  E^{i}(t,{\mathbf{x}})+
\Big[D_j, F^{ji}(t,\mathbf{x})\Big]=\sigma  E^{i}(t,{\mathbf{x}})+ 
\zeta^{i}(t,{\mathbf{x}})~,
\label{eqn:EffThDietrich}
\end{equation}
where $\sigma$ is the color conductivity and $\zeta^{i}(t,\mathbf{x})$ are the stochastic 
color-force fields which satisfy the fluctuation-dissipation relation,
\begin{equation}
  \langle\zeta^{ic}(t_1,\mathbf{x_1})\zeta^{jb}(t_2,\mathbf{x_2})\rangle=2T\sigma\delta^{ij}\delta^{cb}\delta^{3}(\mathbf{x_1-x_2})\delta(t_1-t_2).  
\end{equation}
Due to the non-perturbative nature of this effective theory, we solve Eq.~\ref{eqn:EffThDietrich}
numerically by discretizing it on a three-dimensional lattice with spacing $a_s$ and $N$ sites along 
each spatial direction. The time evolution is also computed numerically using a symplectic integrator.
During the time evolution of these soft color fields, the noise term represents occasional 
kicks received from the thermal bath comprising of hard modes, which are simultaneously dampened due to 
the color conductivity $\sigma$ of the hard gluons. Implementing the fluctuation-dissipation theorem ensures 
that after sufficiently long evolution, the algorithm thermalizes, and color electric fields and gauge links 
are sampled according to a thermal distribution.

We then take a sample thermal configuration consisting of gauge links $U(t,\mathbf{x})$ 
and electric fields $E(t,\mathbf{x})$, evolve them using the Yang-Mills Hamiltonian, and allow fermion 
fields $\Psi(t,\mathbf{x})$ representing a heavy quark of mass $M$ to evolve in this gauge background 
according to the Dirac equation,
\begin{equation}
i\partial_t \Psi(t,\mathbf{x})=\gamma_0\Big[-i\slashed{D}(U)-M\Big]\Psi(t,\mathbf{x})~.
\label{eqn:DiracEqF}
\end{equation}
where $\slashed{D}(U)$ represents the three-dimensional spatial Dirac operator in the presence of 
gauge fields. The fields at time $t$ in terms of the creation (annihilation) operators 
$d^\dagger_{\lambda}(t,\ps^\prime)(b_{\lambda}(t,\ps^\prime))$ for antiquarks (quarks) respectively, at $t=0$ 
can be defined as,

\begin{eqnarray}
        \Psi(t,\xs)= \frac{1}{\sqrt{V}} \sum_{\lambda, \ps^\prime} \Big[  \phi^u_{\lambda,\ps^\prime}(t,\xs) b_{\lambda}(t=0,\ps^\prime) + \nonumber \\ \phi^v_{\lambda,\ps^\prime}(t,\xs) d^{\dagger}_{\lambda}(t=0,\ps^\prime)  \Big]~.
\end{eqnarray}
Here we use the usual box normalization where $V$ represents the three-dimensional volume of the lattice.
The Dirac equation given in Eq.~\ref{eqn:DiracEqF} can be equivalently written in terms 
of the wavefunctions as,
\begin{equation}
i\partial_t \phi(t,\mathbf{x})=\gamma_0\big[-i\slashed{D}(U)-M\big]\phi(t,\mathbf{x})~.
\label{eqn:DiracEqW}
\end{equation}
which can now be solved numerically on the lattice. Performing a Fourier transform, one obtains 
the quark wavefunctions in the momentum space according to the relation, 
\begin{equation}
\Tilde{\phi}^{u,v}_{\lambda,\mathbf{p}}(t,\mathbf{q})=\sum_{\xs} \phi^{u,v}_{\lambda,\mathbf{p}}(t,\xs)
~\rm{e}^{i \mathbf{q}.\xs}~.
\end{equation}
In our simulations, we initiate a free heavy quark (or an anti-quark) in a particular momentum state 
labeled by $\mathbf{p}$ and spin polarization $s$ such that the following commutation relations are 
satisfied,
\begin{eqnarray}
    \langle b_{\lambda}^\dagger(t=0,\ps^{\prime \prime}) b_{\lambda^\prime}(t=0,\ps^\prime) \rangle = \delta_{\lambda \lambda^\prime} \delta_{\lambda^\prime s} \delta(\ps^{\prime \prime}-\ps^\prime) \delta(\ps^\prime-\mathbf{p})&&   \nonumber \\
    \langle d_{\lambda}^\dagger(t=0,\ps^{\prime \prime}) d_{\lambda^\prime}(t=0,\ps^\prime) \rangle  =  0 ~. \nonumber
    \label{eqn:initial_cond}
\end{eqnarray}  
Allowing the quark to evolve in the background of soft gauge fields up to a time $t$, 
the momentum mode occupancy distribution (MMOD) for a quark wavefunction is defined as~\cite{Pandey:2023dzz},
\begin{eqnarray}
    \frac{d^3N}{d^3 \mathbf{q}}(t)= \frac{1}{2N_c} \sum_{\lambda^\prime}
    \big\vert u^\dagger_{\lambda^\prime}(\mathbf{q}) \Tilde{\phi}^u_{s,\mathbf{p}}(t,\mathbf{q})\big\vert^2~.
    \label{eqn:mmod}
\end{eqnarray}
where $u^\dagger_{\lambda^\prime}(\mathbf{q})$ is the non-interacting Dirac momentum eigenstate at 
$t=0$ labeled by momentum $\mathbf{q}$ and a spin+color index $\lambda^{\prime}$. The MMOD, defined in 
Eq.~\ref{eqn:mmod}, represents the number of quark momentum modes $N$ per momentum 
bin between $\mathbf{q}$ and $\mathbf{q}+\delta{\mathbf{q}}$.

When the heavy quark has a finite momentum $\ps$ at $t=0$, its direction sets the longitudinal axis, 
whereas directions orthogonal to it are termed transverse. We can thus calculate the momentum 
distribution functions along the transverse and longitudinal directions separately. The first moment 
of the longitudinal momentum distribution represents the net drag of the heavy quark as a function of time 
$t$. The drag $\eta_D$ and the transverse and longitudinal diffusion coefficients $\kappa_T, \kappa_L$ 
are now functions of the  defined as,
\begin{equation}
    \eta_D(p_L)=-\frac{1}{p_L}\frac{d\langle p_L\rangle}{dt}~,~
    \kappa_{L}=\frac{d\langle (\Delta p_{L})^2\rangle}{dt}~,~
    \kappa_{T}=\frac{1}{2}\frac{d\langle (\Delta p_{T})^2\rangle}{dt}~.
    \label{eqn:DragDiffDef}
\end{equation}
The second moment of the transverse (longitudinal) momentum distribution at each time $t$ 
gives an estimate of the amount of momentum broadening denoted by 
$\Delta p_{T(L)}=p_{T(L)}-\langle p_{T(L)} \rangle$. When second moments increase linearly 
as a function of $t$, it signals diffusive motion of heavy quarks. When the initial 
momentum of the heavy quark is zero, $\kappa_L=\kappa_T$, thereby representing an isotropic 
distribution.

In the limit of infinite heavy quark mass, i.e., $M\to \infty$,  static quarks can be 
represented as Wilson lines. In this case, momentum broadening along the $i$-th spatial direction 
in a non-Abelian gauge theory with $N_c$ colors can be calculated using Kubo's formalism in terms 
of the two-point function of color electric field correlators. Recent studies have focused on 
calculations beyond the infinite mass limit by including the $\mathcal {O}(v/M)$ correction, 
which manifests itself as the two-point correlation function of the color magnetic 
fields~\cite{Bouttefeux:2020ycy}.  The insertion of the color-magnetic fields requires non-trivial 
renormalization.  In our study, we treat heavy quarks as relativistic particles and perform our 
analysis without the need to calculate non-trivial renormalization factors.

\subsection{Coupling heavy quarks to gluons}

In the effective theory, gluons collide with the heavy quark with a frequency $g^4 T$. 
Hence, within a relatively short time scale of $tT\sim 1$, there are $\mathcal{O}(10)$ 
collisions at a typical temperature $\sim 600$ MeV, each imparting a maximum momentum 
of $g^2T/\pi$. The momentum acquired by the heavy quark starting from rest will thus 
acquire the maximum allowed momentum on the lattice within $\mathcal{O}(10)$ collisions for 
reasonably fine lattice spacings. This does not allow for a sufficient diffusive flow of 
the heavy quarks to build up before it saturates; hence, we implement a numerical trick to 
reliably extract the diffusion coefficient, which is described as follows.

While evolving the heavy quark in the background of gauge fields, we scale the coupling of the 
quark to the gauge fields by a factor $\Lambda \ll 1$. This amounts to replacing each gauge link by 
itself but now raised to the power $\Lambda$. In order to keep the Dirac equation invariant, one has 
to scale the heavy quark wavefunctions by $1/\sqrt{\Lambda}$ and momenta by $\Lambda$. Note that 
$\Lambda=1$ gives us the original Dirac equation. Using this technique, the heavy quark mass scales 
by $\Lambda$ such that our simulations are closer to those performed in the infinite physical quark mass 
limit. However, our numerical procedure still allows us to extract the momentum dependence of the heavy 
quark diffusion, which is not possible in the conventional method using the color electric correlators. 
We calculate the momentum distribution of the heavy quark and its $n$-th moments and divide them by 
$\Lambda^n$ respectively, in order to get their physical values.

\section{Numerical details}

\subsection{Lattice parameters}
The $\mathrm{SU(N_c)}$ gauge configurations are generated within the effective theory by discretizing 
the evolution equation of the color electric fields, denoted by Eq.~\ref{eqn:EffThDietrich}, on a 
three-dimensional box with $N=32$ lattice sites along each spatial direction and a lattice 
spacing $a_s$. 
The Langevin time step is chosen to be fine enough such that $a_t/a_s=0.025$. We have performed 
our calculations for both $N_c=2,3$, for a temperature range between $1.6$-$8~T_c$. 
Here $T_c\sim 300$ MeV denotes the deconfinement phase transition temperature, 
expressed in units of the string tension, for both $N_c=2,3$~\cite{Engels:1994xj, Francis:2015lha}. 
The value of $\sigma$ at each temperature is set to its perturbative estimate at 
$\mathcal{O}(g^2)$~\cite{Arnold:1999uy}, 
$\sigma^{-1}(T)=\frac{3N_cg^2T}{4\pi m_D^2}\left[\ln \frac{m_D}{\gamma}+3.041\right],$
where the strong coupling $g$ is determined at scale $2 \pi T$ using the 2-loop $\beta$-function 
with number of quark flavors $N_f=0$. 
The quantity $1/\gamma$ denotes the mean free path of the color-changing processes, which has a 
value of $\gamma/m_D \approx 0.883$-$1.034$ at the temperature range we are studying~\cite{Guin:2025lpy}. 
The Debye mass $m_D$ is calculated at one-loop order such that $m_D=gT\sqrt{\frac{N_c}{3}}$. The 
lattice spacing $a_s$ can be written in physical units by matching the energy density that we 
calculate at each temperature to the Stefan-Boltzmann limit which results in $a_s T= 1.45$. 
However, the continuum extrapolated lattice results for the energy density in $\mathrm{SU(N_c)}$ 
gauge theory deviates from Stefan-Boltzmann value at $T=2~T_c$ by about $\sim 18\%$~\cite{Borsanyi:2012ve} 
which results in an error of about $6\%$ in the determination of the lattice spacing. The quark mass is set to 
a value $a_sM=10.15$ such that we are in the regime where $M\gg T$. We also ensure that the violation 
of Gauss law at each time step is $\sim 10^{-13}$ such that we sample only the physical states in our 
numerical computations.

\subsection{Heavy quark observables}

At $t=0$, the heavy quark is in a single momentum eigenstate with initial MMOD given by
$\delta^3(\mathbf{q}-\mathbf{p})$. We choose the initial momentum along a particular spatial 
direction, e.g., x-direction whose magnitudes are $\vert \ps \vert/T=0, 0.135, 0.271, 0.405, 0.535$, 
$0.658, 0.768, 0.858, 0.923$. The allowed values of momenta on the lattice $\hat{q}_i$ are discrete 
which we implement as 
\begin{eqnarray}
    \hat{q}_i= - \sum_{k} ~ C_k ~\sin \Bigg(k \frac{2\pi n_i}{N}\Bigg)~,
\end{eqnarray}
where $n_i=0,...,N-1$ is the mode index of lattice momenta along the $i$-th spatial 
direction and $\{0,1,..,C_{k-1}\}$ are coefficients used for $\mathcal {O}(a^{k})$ 
tree-level improvement of the Wilson-Dirac operator. We use the NLO improvement consisting 
of $k=2$ coefficients which are $C_1=4/3$ and $C_2=-1/6$. We also explicitly mention the NLO 
improved Wilson-Dirac operator for heavy quarks used in our simulations,  
\begin{eqnarray}
&& (-i\slashed{D}_{W})_{\xs,\ys}  =\sum_{k,i}  \frac{C_k}{2a_s}
\Big[ \Big(-i\gamma^{i}- k \Big) U_{\xs,+ki} \delta_{\xs+k\ib,\ys}  \nonumber \\
&& \qquad\qquad  +2 k~  \delta_{\xs,\ys} - 
\Big(-i \gamma^{i}+k  \Big)U_{\xs,-ki}
\delta_{\xs-k\ib,\ys} \Big] \;,
\end{eqnarray}
where the gauge links $U_{\xs,\pm ki}$ are defined as
$U_{\xs,+ki}=\prod_{n=0}^{k-1} U_{i,\xs+ni}\;,~U_{\xs,-ki}= \prod_{n=1}^{k} U^{\dagger}_{i,\xs-ni}\;.$
The quark wave functions are evolved quantum-mechanically as a function of time in the background of 
classical color gauge fields using a leapfrog integrator with a time step $a_t=0.025 a_s$. These quark wave 
functions at different times $t$ are projected onto non-interacting momentum eigenstates labeled by momentum 
values $\hat{p}_i$ allowed on a lattice according to Eq.~\ref{eqn:mmod} in order to calculate the momentum 
mode occupancy distribution (MMOD). The second moment of the distribution quantifies the momentum 
broadening as a function of $t$ due to collisions, which is calculated on the lattice  
with soft gluons 
\begin{eqnarray}
    \langle {\hat{p}_i}^2 \rangle (t) ~ = \sum_{\{n_i\} }\Delta \hat{p}_i \times {\hat{p}_i}^2 \times ~\sum_{\{n_j\} } \Bigg( \frac{d^3 N}{d^3 \mathbf{p}} ~  \prod_{j \neq i} \Delta \hat{p}_j   \Bigg)~.
    \label{eqn:LatDefqsq}
\end{eqnarray}
where $\{n_i\}$ represents the momentum mode along the $i$-th spatial direction and $j$ represent all other directions 
orthogonal to $i$.

\section{Results}

\subsection{\texorpdfstring{Heavy quark momentum diffusion for $M\gg T,~ \vert \ps\vert=0$}{Heavy quark momentum diffusion for M>>T}}

\begin{figure}[h]
    \centering
    \includegraphics[width=0.45\textwidth]{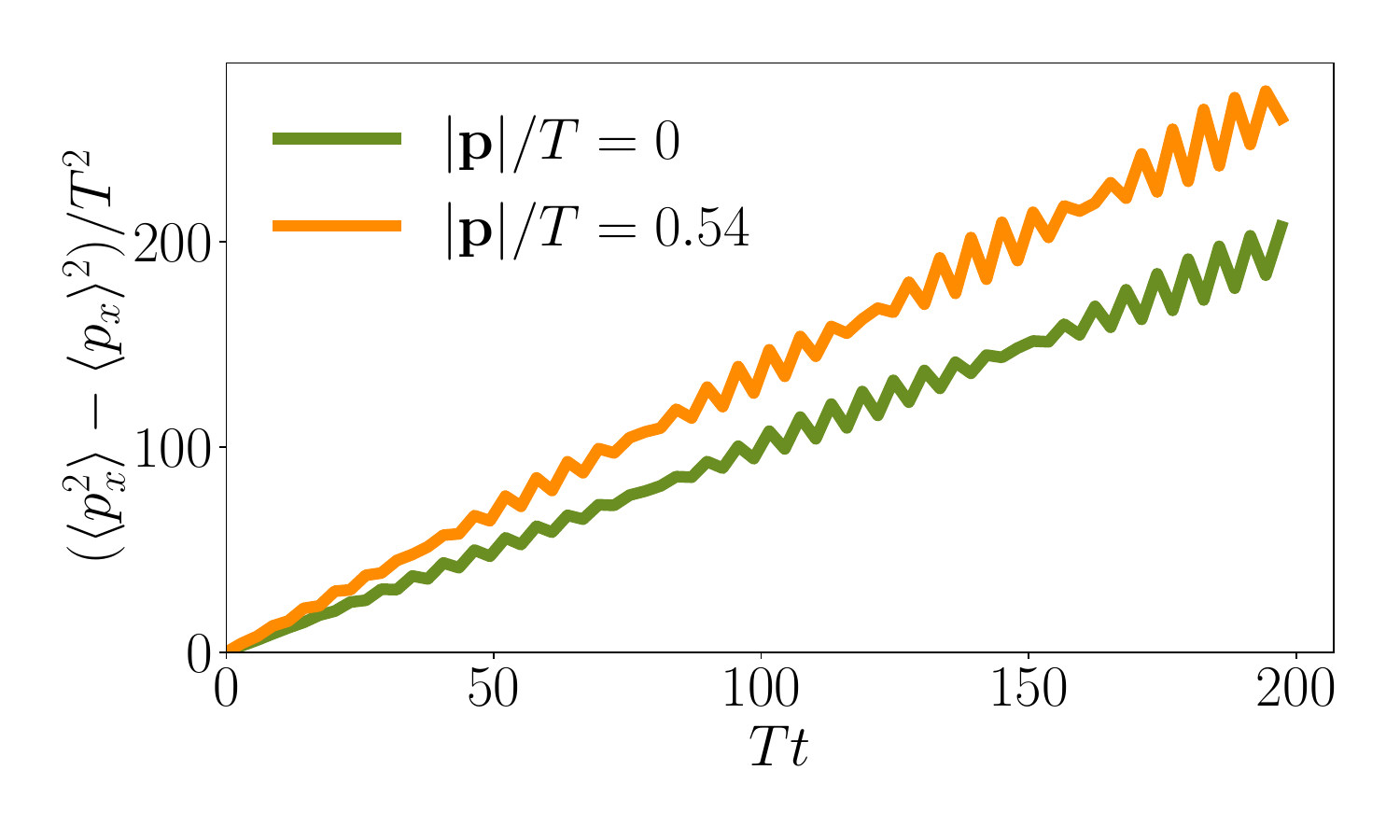}
    \caption{Variance of the momentum distribution $(\langle p_x^2\rangle - \langle p_x\rangle^2 )/T^2$ for a heavy quark with a zero and finite initial momenta $|\mathbf{p}|/T=0.54$ in a typical thermal configuration consisting of 
    non-perturbatively interacting soft SU(3) gluons at $T\sim 480$ MeV. }
    \label{fig:fermionic_px2_2Tc_w_Lambda}
\end{figure}

We calculate the variance of the momentum distribution along the $x$-direction for a heavy quark 
with a zero initial momentum by solving the Dirac equation, but by scaling the quark coupling to 
gauge fields by a factor $\Lambda=0.005$. We first focus on a particular temperature which is 
$T \simeq 1.6 ~T_c$ and $N_c=3$. The second moment of the MMOD as a function of 
time $Tt$ is shown in Fig.~\ref{fig:fermionic_px2_2Tc_w_Lambda}. The late-time growth of the 
variance is linear as a function of time, indicative of a diffusive motion. There are tiny 
fluctuations about the linear rising contribution, occurring due to the heavy but finite mass 
of the quark on the lattice. The diffusive motion is also visible for all values of initial 
momentum we have considered in this work. As a representative, we also show how the 
variance of the momentum distribution varies as a function of $Tt$ for an initial momentum 
$\vert \ps\vert=0.54~T$ in Fig.~\ref{fig:fermionic_px2_2Tc_w_Lambda}. At this temperature, 
we extract the value of $\kappa/T^3= 1.00(5)$ for a fit to the data of momentum variance in 
the range $Tt=100$-$200$ for $\vert \ps\vert=0$. Our results are consistent with the 
values of $\kappa/T^3 \approx 1.5$-$2.8$~\cite{Banerjee:2022gen}, $\approx 1.31$-$3.64$
~\cite{Brambilla:2020siz} calculated in SU(3) gauge theory at $1.5~T_c$. However, the errors
in our data are comparatively smaller. The momentum diffusion caused by these 
strongly interacting soft gluons can also account for the values of $\kappa/T^3 \sim 1.1$-$4.6$ 
reported for bottom quarks obtained in 2+1 flavor QCD at $T \sim 444$-$500$ MeV~\cite{HotQCD:2025fbd}. 
These observations suggest that the diffusion of heavy quarks at high temperatures is predominantly 
driven by the kicks received from non-perturbatively interacting soft gluons.

We will end this section with two more interesting observations. At early times 
$T t<20$, the $\langle p_x^2 \rangle$ for heavy quarks with zero initial momentum 
grows quadratically as a function of time, eventually entering into a diffusive regime 
at late times where $\langle p_x^2 \rangle \propto t $. The onset of the diffusive regime 
is comparatively faster for higher values of the initial momenta of heavy quarks. This might 
be a signature that slower heavy quarks undergo collective motion in the medium consisting of 
non-perturbatively interacting soft gluons. Second, the values of $\langle p_x^2 \rangle$ extracted 
from the color electric two-point correlators within the effective theory do not show a diffusive 
growth in the time interval where we have extracted $\kappa$ from MMOD of heavy quarks. This is 
possibly due to the fact that hard static gluons are not present within the effective theory,
which otherwise mimics heavy quarks in the full gauge theory. Gluons with the highest possible allowed 
momenta $\lesssim gT$ will still participate in a collective rather than a diffusive motion with 
the thermal plasma where gluons have typical momenta $< g^2 T/\pi$.

\subsection{Temperature dependence of $\kappa$ at $\vert \ps \vert=0$}

\begin{figure}[h!]
    \centering
    \includegraphics[width=0.45\textwidth]{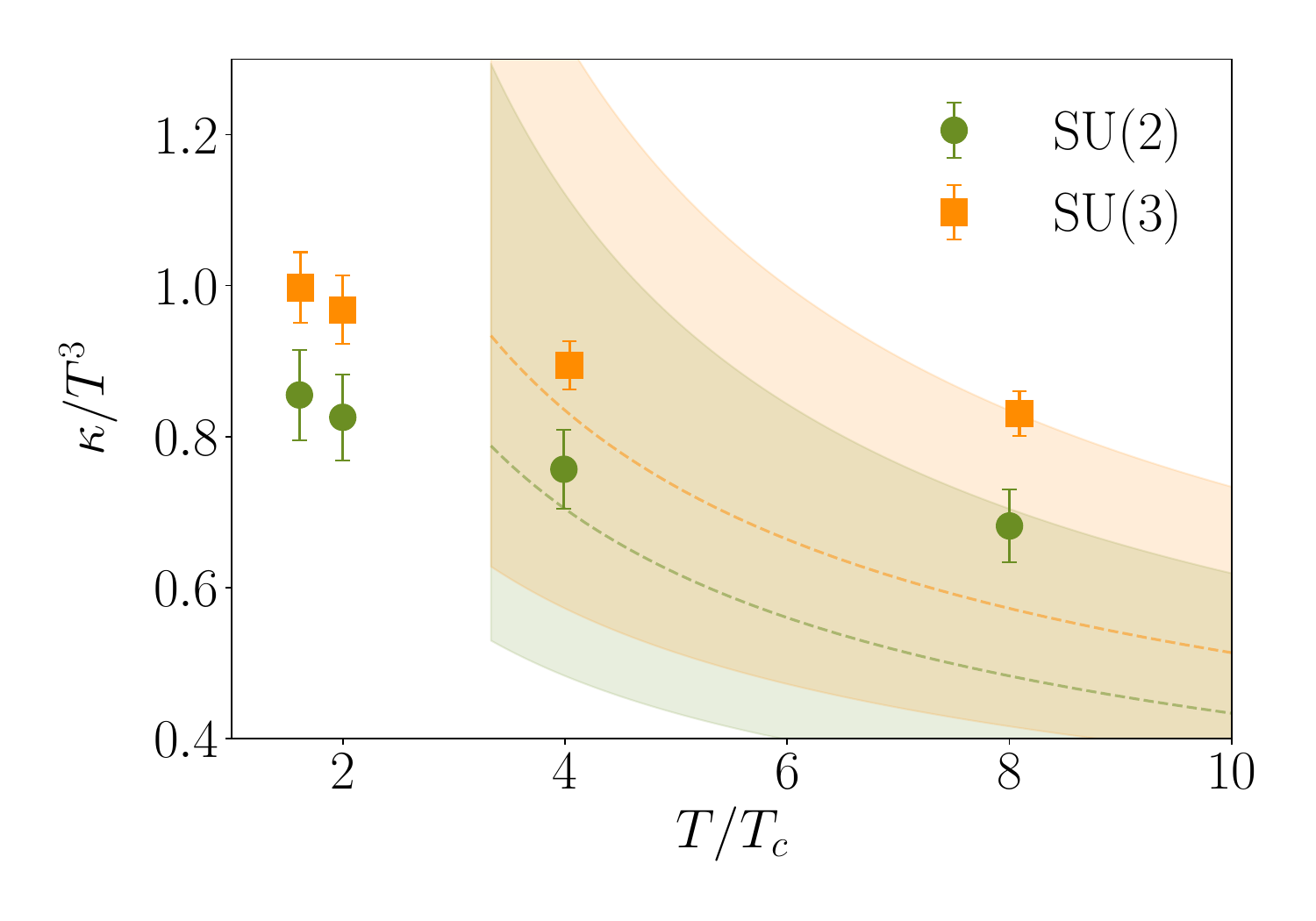}
    \caption{The $\kappa/T^{3}$ for zero initial momentum of a heavy quark, obtained in an effective theory of soft (green points) SU(2) and (orange points) SU(3) gluons, shown as a function of temperature. For comparison, we show the values of the same ratio in 4D QCD without dynamical quarks as a function of temperature, where the parametric dependence of $\kappa$ was taken from Ref.~\cite{Brambilla:2020siz}.}
    \label{fig:fermionic_kappa_vs_T_p0}
\end{figure}

We next calculate $\kappa$ for heavy quarks with zero initial momentum $\vert \ps \vert=0$,  
for a wide range of temperatures within the thermal effective theory for $N_c=2,3$.  
The results for $\kappa/T^3$ are compiled in Fig.~\ref{fig:fermionic_kappa_vs_T_p0}. We also compare 
our results with the parametrization of $\kappa(\vert \ps \vert=0)$ as a function of $T$ 
obtained in 4D SU(3) gauge theory Ref.~\cite{Brambilla:2020siz}, shown as the dotted line 
within a band in the same figure, where the coupling is determined at the scale $2\pi T$. 
The upper and lower boundaries of the band correspond to the choices of the scale $\pi T$ 
and $4\pi T$ respectively,  at which the strong coupling is calculated. It is worth 
mentioning that the parametrization of $\kappa$ we are comparing with has the same 
functional dependence as obtained from NLO HTL perturbation theory~\cite{Caron-Huot:2007rwy} 
except that the coefficient of the term $m_D/T$ was larger by $64\%$~\cite{Brambilla:2020siz}. 
We observe a good agreement within the large-scale uncertainties inherently present in the 
parametrization; however, the errors in our data are smaller compared to the results for 
$\kappa$ obtained in 4D SU(3) gauge theory. However, we also note that $\kappa/T^3$ decreases gradually with 
increasing temperatures, which cannot be accounted for e.g. within HTL perturbation theory 
at NLO since the broadening due to multiple soft scatterings is not accounted for completely 
within HTL.

\subsection{Drag and diffusion at $T=1.6~T_c$}

\begin{figure}[h!]
    \centering
    {\includegraphics[width=0.45\textwidth]{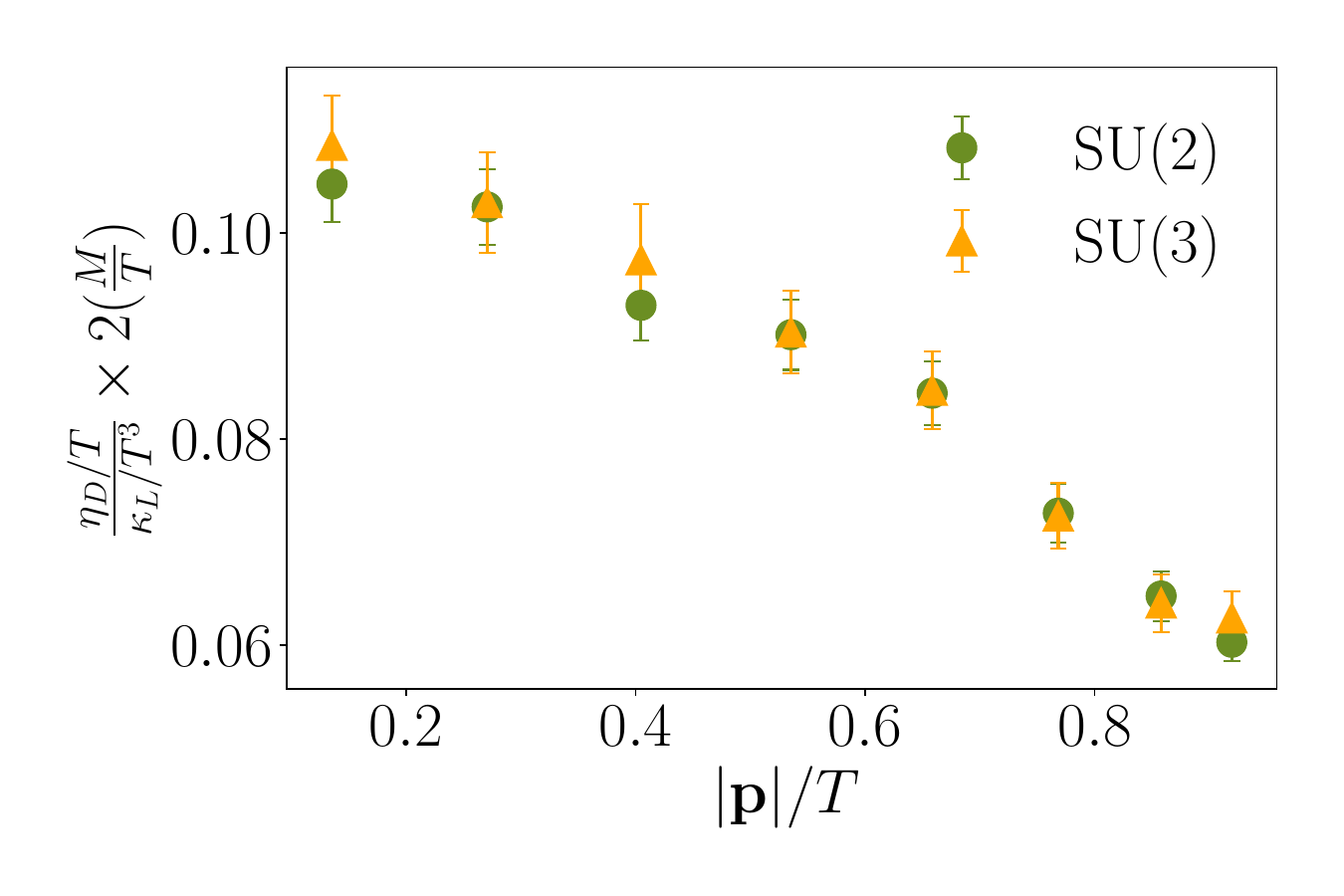}} 
    \caption{The ratio of drag and diffusion coefficients $\frac{\eta_D/T}{\kappa_L/T^3}\times2\Big(\frac{M}{T}\Big)$ as a function of initial quark momenta $|\ps|/T$, for a high temperature effective theory of soft (green) SU(2) and (orange) SU(3) gluons at $T=1.6~T_c$.}
    \label{fig:fermionic_eta_over_kappa_vs_p_600MeV}
\end{figure}

At $T=1.6~T_c$, we calculate the MMOD for heavy quarks that have an initial momentum, both along 
and transverse to their direction. This allows us to extract the drag and diffusion coefficients 
from the MMOD in the diffusive regime, along the longitudinal direction, using Eq.~\ref{eqn:DragDiffDef}. 
The results for the  ratio  $\frac{2M}{T}\frac{\eta_D/T}{\kappa_L/T^3}$ as a function of $\vert \ps \vert/T$ 
for both $N_c=2,3$ are shown in Fig.~\ref{fig:fermionic_eta_over_kappa_vs_p_600MeV}. Even for the smallest  
value of $\vert \ps \vert/T$, which corresponds to a velocity of the heavy quark $v \sim 0.02$, there is a significant 
deviation of the ratio from unity, which is realized when the Einstein relation is applicable. Our lattice data for 
the ratio does not match its perturbative value of $1-5T/M=2/7$ when the next-to-leading order velocity 
correction~\cite{Bouttefeux:2020ycy} is considered. This observation is not sensitive to the 
number of colors.

Incidentally, the Einstein relation is known to hold only up to leading logarithm accuracy in perturbative 
QCD when the velocity of the heavy quark $v>0$~\cite{Moore:2004tg}. However, in strongly 
coupled $\mathcal{N}=4$ supersymmetric Yang-Mills theory, the Einstein relation is badly broken
~\cite{Gubser:2006nz}. Recently, generalization of the Einstein relation to quantum field theories 
with non-Gaussian fluctuations has been discussed~\cite{Rajagopal:2025rxr,Rajagopal:2026urq}, where 
the need to quantify higher moments of the MMOD for heavy quarks is emphasized. Our approach allows us to 
follow upon this suggestion, in particular, we indeed observe a non-zero but noisy third moment for MMOD 
at the lowest momentum of heavy quarks. Our results for the moments of MMOD are much closer to 
expectations from a strongly coupled theory, as expected, since we are specifically studying the 
non-perturbative effects due to soft gluons. Our study thus provides a non-trivial demonstration of 
a significant deviation from the Einstein relation in a strongly interacting non-conformal system.

\subsection{Momentum dependence of heavy quark diffusion coefficients }

\begin{figure}[h!]
    \centering
    {\includegraphics[width=0.45\textwidth]{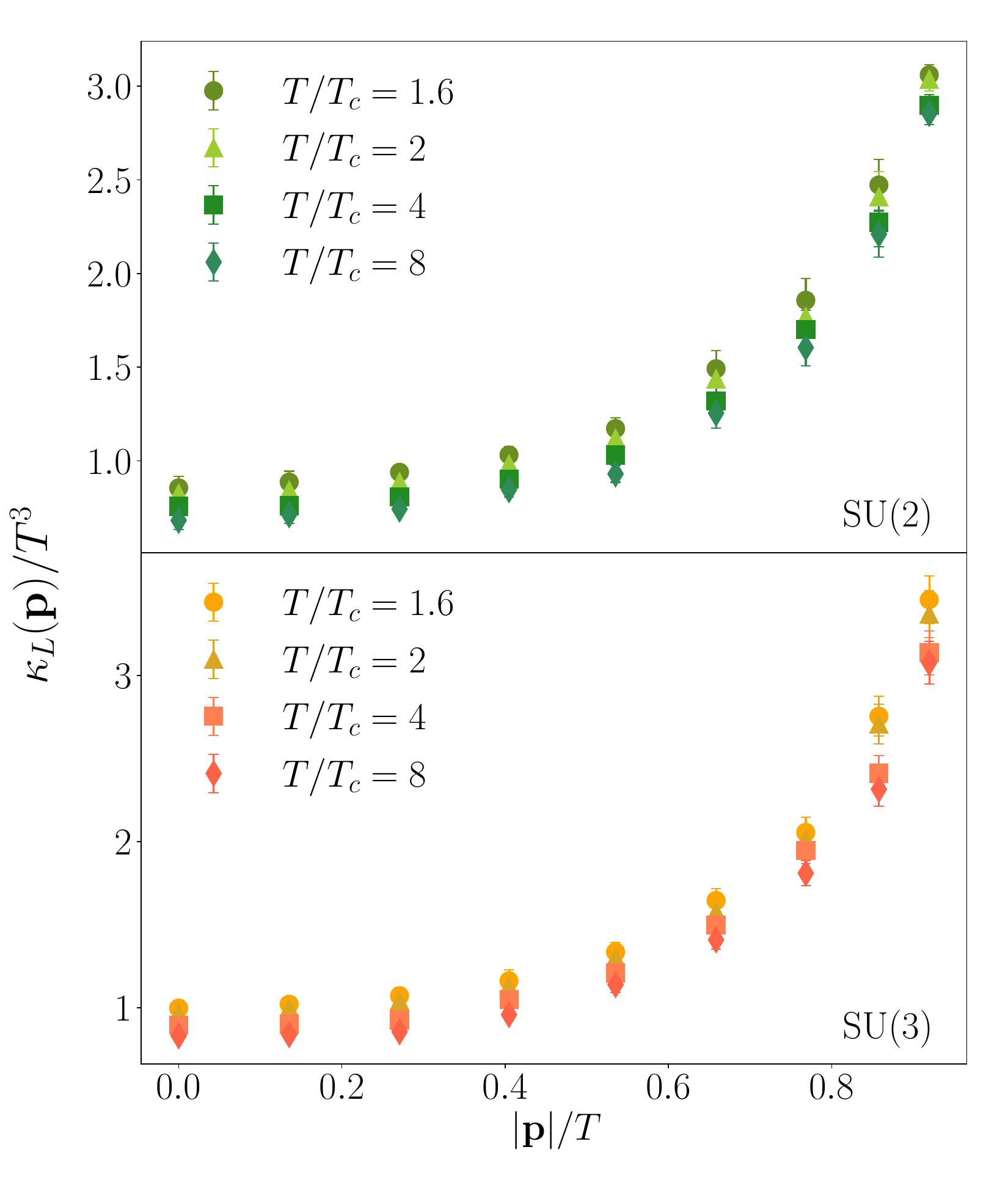}}
    \caption{The $\kappa_L(\ps)/T^{3}$ of a heavy quark  as a function of its initial momenta 
    $|\ps|/T$ for different temperatures in a (top panel) SU(2) and (bottom panel) 
    SU(3) plasma consisting of soft gluons.}
    \label{fig:fermionic_kappa_L_vs_p_diffT}
\end{figure}

\begin{figure}[h!]
    \centering
    {\includegraphics[width=0.45\textwidth]{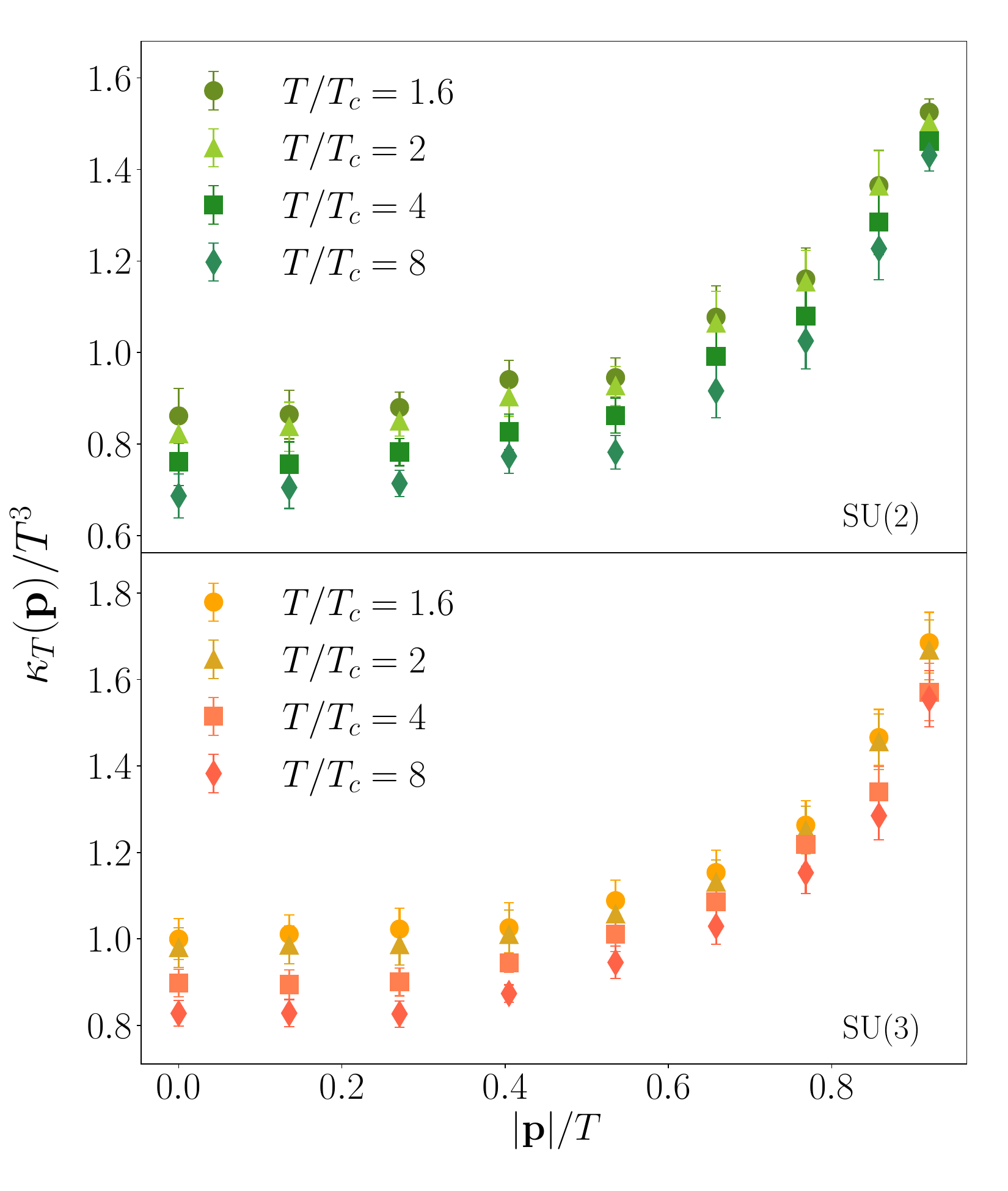}}
    \caption{The $\kappa_T(\ps)/T^{3}$ of a heavy quark as a function of its initial momenta 
    $|\ps|/T$ for different temperatures in a (top panel) SU(2) and (bottom panel) 
    SU(3) plasma consisting of soft gluons.}
    \label{fig:fermionic_kappa_T_vs_p_diffT}
\end{figure}

An important advantage of our formalism over the existing lattice computations performed close 
to the infinite quark mass limit is that it can be used to calculate the momentum dependence 
of diffusion coefficients. The longitudinal and transverse momentum diffusion coefficients, 
scaled by $T^3$, are shown as a function of the initial momentum $\vert \mathbf{p} \vert$
of a heavy quark in Figs.~\ref{fig:fermionic_kappa_L_vs_p_diffT} and 
\ref{fig:fermionic_kappa_T_vs_p_diffT} respectively. Both these coefficients increase with 
increasing $\vert \mathbf{p} \vert/T$; however, the diffusion is more along the longitudinal 
direction. Moreover, the magnitude of diffusion coefficients decreases with increasing temperatures 
at each value of $\vert \mathbf{p} \vert$. These trends visible in the transport coefficients are 
consistent with those obtained within kinetic theory~\cite{Moore:2004tg} at leading log order in 
$m_D$ and using the HTL propagator for gluons.

However, a direct quantitative comparison of our lattice results for transport coefficients with those 
from the Boltzmann equation is not very well motivated. The validity of Boltzmann transport is limited 
to weakly coupled plasmas that have quasiparticle excitations such that $m_D \ll T$. Moreover, the 
choice of a suitable Boltzmann collision kernel typically involves further model-dependent assumptions. 
These factors all make it difficult to relate the Boltzmann collision kernel to transport properties 
calculated from first principles in a non-perturbative thermal field theory.

\section{Implications of our findings \& Outlook}

In this work, we have developed a framework based on first-principles lattice gauge theory to calculate the momentum dependence of the motion of a heavy quark inside a non-Abelian plasma consisting of non-perturbatively interacting soft gluons. Treating heavy quarks as Dirac particles allows us to calculate the momentum mode occupancy distribution (MMOD) and its moments as a function of time for different values of their initial momenta, even when $M\gg T$.  This allows us to extract the ratio of the drag and diffusion coefficients for the first time with a very good precision, which indicates deviation from the Einstein relation at $\vert \ps\vert/T \sim 0.1$ at $1.6~T_c$. The deviation is even more drastic for faster-moving heavy quarks. In order to ensure kinetic equilibration for heavy quarks with arbitrary initial velocities, the Boltzmann transport formulation has been very widely studied and used in recent years to benchmark earlier Langevin treatments~\cite{Das:2013kea,Cao:2018ews,Xu:2017obm,Ke:2018tsh}.  Our study provides a complementary non-perturbative method to extract the momentum and temperature dependence of the heavy quark transport coefficients beyond the usual limitations in the Boltzmann approach.

 Interestingly, our results for $\kappa(\vert \ps\vert)$, both along the longitudinal as well as transverse to its initial direction of motion, exhibit a strong dependence on the initial momentum carried by the heavy quark. Such a significant momentum dependence cannot be explained within the Boltzmann approach and thus has implications for determining the time required for heavy quarks to achieve kinetic equilibration in QGP. We anticipate that our results will contribute towards a more realistic modeling of heavy quark dynamics~\cite{He:2011qa,Rapp:2018qla} in a heavy-ion collision event. For example, the $\kappa(\vert \ps\vert)$ that we report here can reduce systematic uncertainties in the extraction of elliptic flow for heavy quarks, or the temperature dependence of the drag $\eta_D$ can play a role in simultaneously explaining the magnitude of $R_{AA}$ and $v_2$~\cite{Das:2015ana} observed in heavy-ion collision experiments.

At present, our framework can account for the momentum dependence of heavy quark diffusion in the limit $M\gg T$.  In the future, we would like to generalize our formalism to study, in addition, the flavor dependence of heavy quark dynamics in QGP for a temperature regime relevant for heavy-ion collider experiments.

\section*{Acknowledgements}

We thank Sayak Guin for his help in generating the gauge configurations and for valuable discussions. 
We are also grateful to S\"{o}ren Schlichting for discussions on a related project.
We acknowledge the HPC center at the Institute of Mathematical Sciences, where simulations were performed.
This research was supported in part by the International Centre for Theoretical Sciences (ICTS) for participating 
in the program--Hard probes in non-equilibrium QCD matter, 2026 (ICTS/NQCD2026/03).

\bibliographystyle{apsrev4-2}

\bibliography{FiniteTHQPaper_Harshit}

\end{document}